\documentclass[preprint2]{aastex}
\usepackage{graphicx}                    

\usepackage{color}                       
\usepackage{url}                         

\newcommand{\et}{{\it et al.}}

\shortauthors{Mackay et al.}
\shorttitle{Dispersal of an Active Region}

\begin{document}

\title{Modeling the Dispersal of an Active Region: Quantifying Energy Input into the
Corona}

\author{Duncan H Mackay$^1$, L M Green$^2$ and Aad van Ballegooijen$^3$}

\affil{$^1$ School of Mathematics and Statistics, University of St Andrews, North Haugh, 
St Andrews, Fife, Scotland, KY16 9SS
\\
$^2$ University College London, Mullard Space Science Laboratory, Holmbury St. Mary, 
Dorking, Surrey, RH5 6NT, UK \\ 
$^3$ Harvard-Smithsonian Center for Astrophysics, 60 Garden Street, Cambridge, MA 02138}
 
\email{duncan@mcs.st-and.ac.uk,lmg@mssl.ucl.ac.uk, vanballe@cfa.harvard.edu}

\begin{abstract}
In this paper a new technique for modeling non-linear force-free fields directly from line of sight
magnetogram observations is presented. The technique uses sequences of magnetograms directly as lower boundary 
conditions to drive the evolution of coronal magnetic fields between successive force-free equilibria
over long periods of time. It is illustrated by applying it to MDI 
observations of a decaying active region, NOAA  AR 8005. The active region is modeled during a 4 day period 
around its central meridian passage. Over this time, the dispersal of the active region is dominated 
by random motions due to small scale convective cells. Through studying the build up of magnetic energy in
the model, it is found that such small scale motions may inject anywhere from $2.5-3 \times 10^{25}$ ergs 
s$^{-1}$ of free magnetic energy into the coronal field. Most of this energy is stored within the center
of the active region in the low corona, below 30 Mm. After 4 days 
the build-up of free energy is 10$\%$ that of the corresponding potential field. This energy buildup, is 
sufficient to explain the radiative losses at coronal temperatures within the active region.  
Small scale convective motions therefore play an integral part in the energy balance of the corona. This
new technique has wide ranging applications with the new high resolution, high cadence observations from 
the SDO:HMI and SDO:AIA instruments.
\end{abstract}


\maketitle

\section{Introduction}

The solar corona is a complex environment, where much of its complexity is due to magnetic fields. 
These magnetic fields are produced through a dynamo action near the base of the convection zone
\citep{charb05}. Once formed they may become buoyantly unstable, rise through the 
convective zone and break through the photosphere \citep{arch04,mag04,gals07,murr08,fan09}
where they are observed as either sunspots in white light
or active regions in magnetograms. Such active regions structure the Sun's atmosphere and provide energy 
for eruptive phenomena such as solar flares \citep{benz08} and CMEs \citep{crem06}. 

Once active regions form, small scale motions such as granular and supergranular flows result in 
the decay of the active region and the dispersal of magnetic flux across the solar surface in a
random walk \citep{leigh64}. This random walk leads to the convergence and cancellation of magnetic flux along 
Polarity Inversion Lines (PILs), in addition to the spreading of magnetic fields. Such evolution
of magnetic elements in the photosphere acts as a driver for the build up of free magnetic energy 
in the solar corona. The subsequent evolution of the coronal field through quasi-static
equilibria, in response to these motions, may be mathematically modeled through force-free magnetic 
fields, magnetic fields which 
satisfy $\bf{j} \times \bf{B} = 0$ where ${\bf j} = \alpha {\bf B}$ \citep{priest82}.
  
In recent years, non-linear force-free field modeling has received much attention \citep{sch06,met08}. A
non-linear force-free field, is a special class of force-free field, where the scalar function
$\alpha(r)$ is a function of position, but must be constant along any field line. Non-linear
force-free fields are of particular interest as they may contain free magnetic energy
\citep{wolt58}. The construction of non-linear force-free fields (NLFFF) may be
broadly split into two groups: static and time-dependent models.
Examples of static modeling are the ``extrapolation" of photospheric
vector magnetic fields into the corona \citep{sch06,reg07,derosa09,wheat09,jing10}, and direct 
fitting of NLFFF models to observed filaments and coronal structures 
\citep{bal04,bob08,su09b,sav09}.  Static modeling involves constructing either individual 
NLFFFs or independent sequences of magnetic configurations, where there is no direct correlation 
or evolution between the different fields.

In contrast, time-dependent quasi-static modeling evolves the coronal magnetic field through 
continuous sequences of related non-linear force-free fields based on the evolution of a continuous
time dependent lower boundary condition. This boundary condition may be specified through
either idealized magnetic field configurations \citep{bal00,mac03,mac05,mac06a,mac09} or from 
observations \citep{mac00,yea07,yea08}. Only through using 
dynamic models can the build up of free magnetic energy and magnetic helicity in the corona be 
studied, along with the evolution of isolated flux and current systems. Such a technique has been 
successfully applied in the past to model the evolution of the global corona
\citep{mac06a,mac06b,yea09}, determine the origin and evolution of solar filaments
\citep{yea08,mac09}, study the formation and lift off of magnetic flux ropes 
\citep{mac06b,yea09,yea10a} and finally the origin of the Sun's Open Magnetic Flux 
\citep{mac06b,yea10b}.

In this paper we apply time dependent quasi-static non-linear force-free modeling to model the
effect that the dispersal of an active region has on the overlying coronal magnetic field. 
A key difference between this paper and previous studies is the manner
in which the time dependent lower boundary condition is applied. In previous studies either
idealized magnetic field distributions or simplified surface configurations derived
from observations were prescribed. In the present study, a new technique of prescribing the time 
dependent lower boundary condition is presented. The technique allows the use of observed 
line of sight component magnetograms directly as lower boundary conditions, reproducing as observed, 
the dispersal of the photospheric flux of the active region. From this the 
effect of the surface motions on the coronal magnetic field and the subsequent
energy input into the corona is determined. 

The active region chosen to illustrate this new technique is NOAA AR 8005. It is chosen as it
was an isolated region and during the period of observations there appears to be no significant
flux emergence
and the flux is well balanced throughout. The structure of the paper is as follows. In Section 
2 the main properties of the decaying active region are discussed. In Section 3 the 
modeling technique for both the photospheric and coronal fields are described. The results 
of the simulation are given in Section 4, while in Section 5 some simple calculations
are compared to the simulations to verify the results. Finally a discussion and conclusions 
are given in Section 6.

\begin{figure}[t]
\centering\includegraphics[scale=0.6]{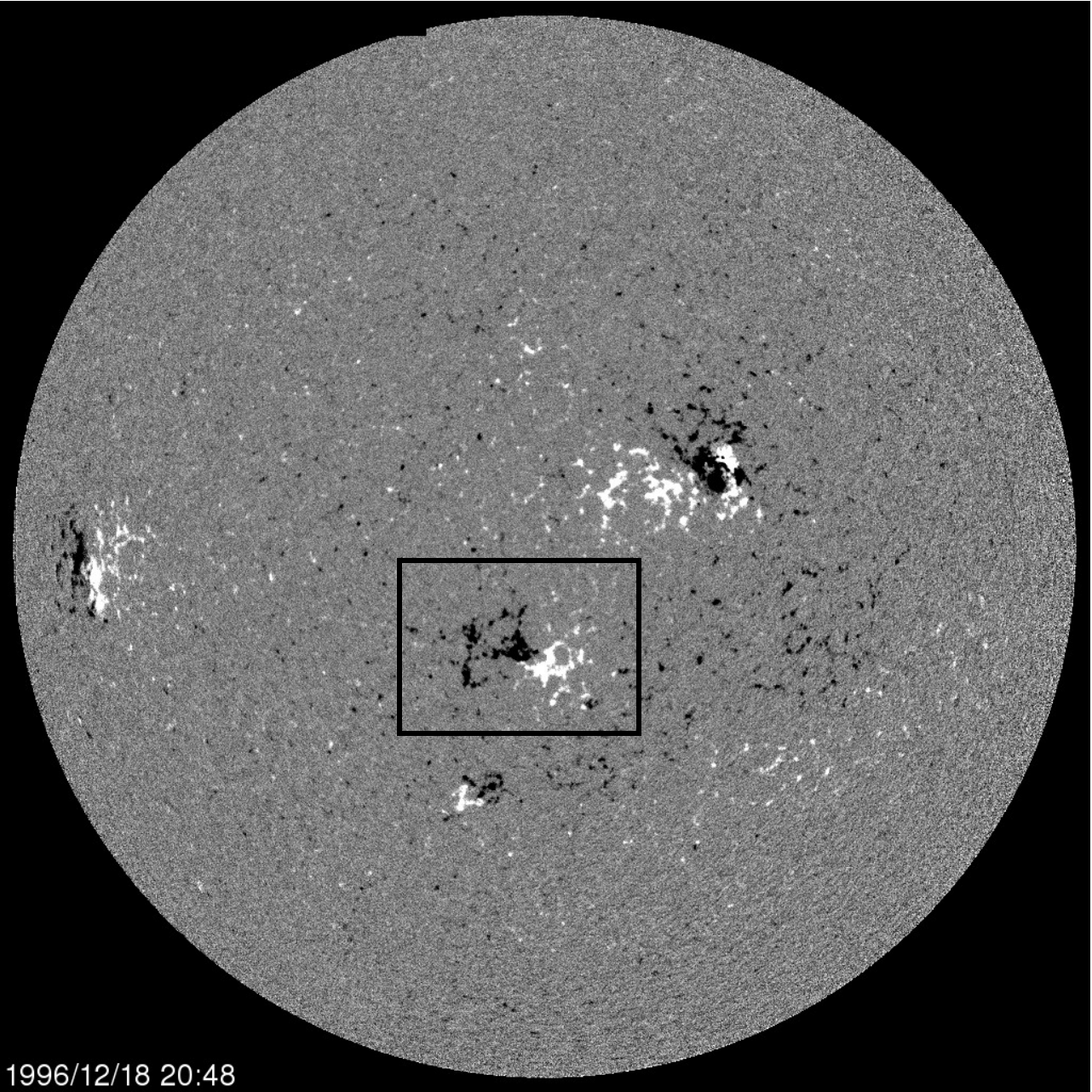}
\caption{SoHO MDI $96$ min line of sight component magnetogram for 18$^{th}$ Dec 1996 where white 
represents positive magnetic polarities and black negative magnetic polarities. The active 
region of interest, NOAA 8055, lies in the southern hemisphere at central meridian and is 
enclosed by the black box.}
\label{fig:fig1}
\end{figure}

\section{Observations}

The decaying active region considered in this study is NOAA AR 8005. It emerged
on the far side of the Sun and as it rotated onto the limb only dispersed magnetic polarities,
without sunspots were present. In Figure~\ref{fig:fig1} a full disk SOHO:MDI \citep{Scherrer95}
$96$min line of sight 
component magnetogram from 18$^{th}$ Dec 1996 can be seen. The active region of interest 
lies at central meridian just below the equator. It has a simple magnetic morphology with a 
single positive and negative polarity. 

\begin{figure}[t]
\centering\includegraphics[scale=0.6]{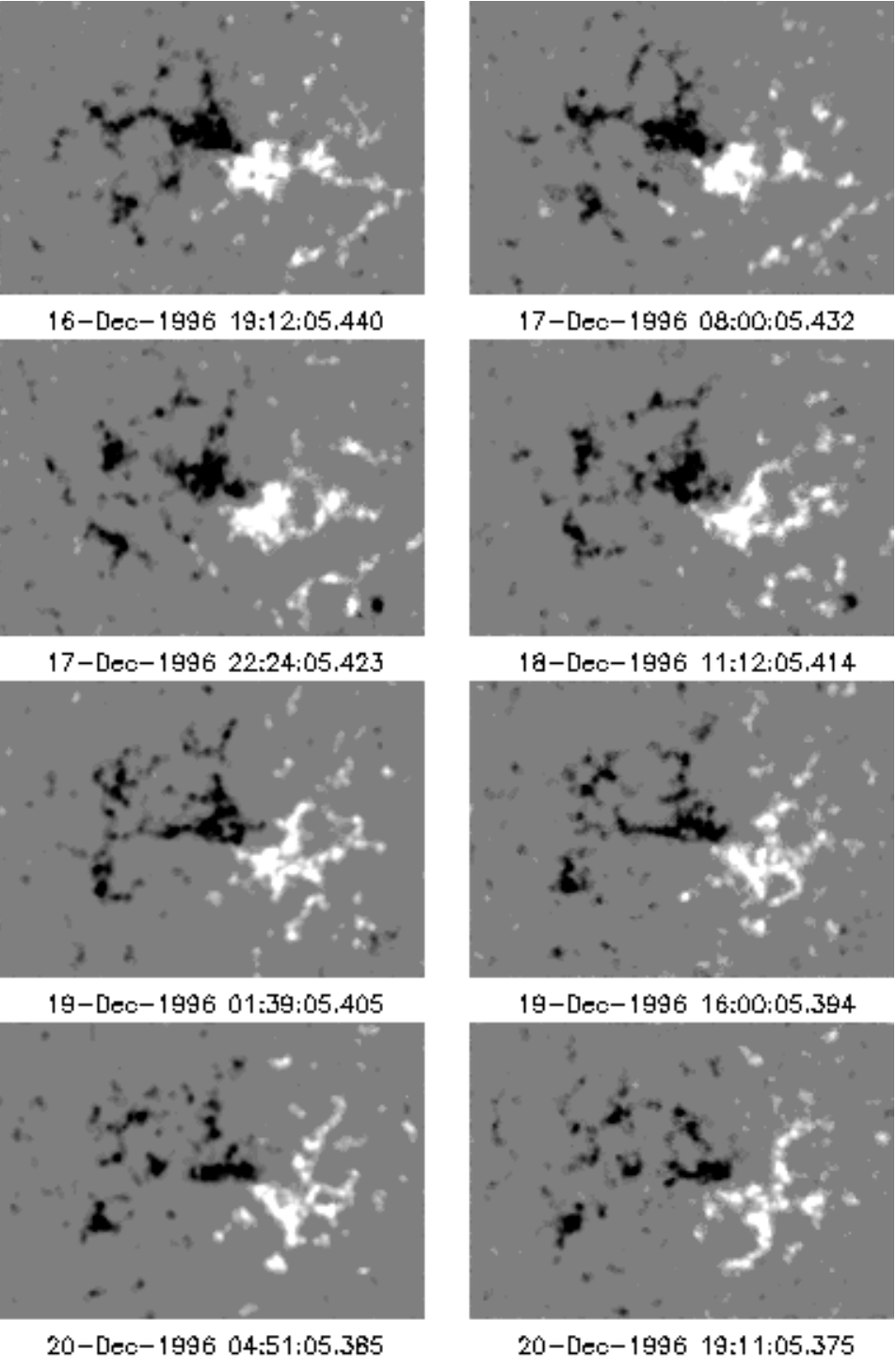}
\caption{Time sequence of de-rotated SoHO MDI $96$ min line of sight component magnetograms  
of NOAA 8005. The time sequence covers the evolution of the active region 2 days before
and 2 days after central meridian passage (18$^{th}$ Dec 1996) where white represents positive 
magnetic polarities and black negative magnetic polarities. The images show the active
region after it has been corrected for flux balance.}
\label{fig:fig2}
\end{figure}

To study the evolution of the active region, SoHO:MDI $96$min magnetograms\footnote{To produce 
the $96$ min magnetograms, 5 individual SoHO/MDI magnetograms of higher time cadence and noise error 
of $\pm$20G per pixel are averaged. Correspondingly the resulting $96$ min magnetograms have a lower 
noise error per pixel of $\pm$9G. } are obtained from
19.12.05 UT on 16$^{th}$ December 1996 to 20.47.05 UT on 20$^{th}$ December 1996, spanning a
period of four days. The four day period was chosen to include, two days before and two days 
after central meridian passage, which occurs late on 18$^{th}$ December. During this period 
as LOS projection effects are minimized, the measurements of the magnetic flux are 
known to be the most reliable. 
In total 61, $96$ min magnetograms cover the
period of interest. The data are corrected for the area foreshortening 
that occurs away from central meridian using the IDL Solar Software routine {\it drot\_map}.
An area of $181 \times 126$ pixels is cut
out of the rotated magnetograms centered on NOAA 8055 where each pixel is $1.977''$.
A time sequence of these can be seen in Figure~\ref{fig:fig2}. The area was chosen to be large 
enough to encompass the whole active region, but small 
enough that approximate flux balance is achieved. The active regions that lie to the north and 
south of NOAA 8005 have no contribution to the flux in the area considered. 

\begin{figure}[t]
\rotatebox{270}{\centering\includegraphics[scale=0.4]{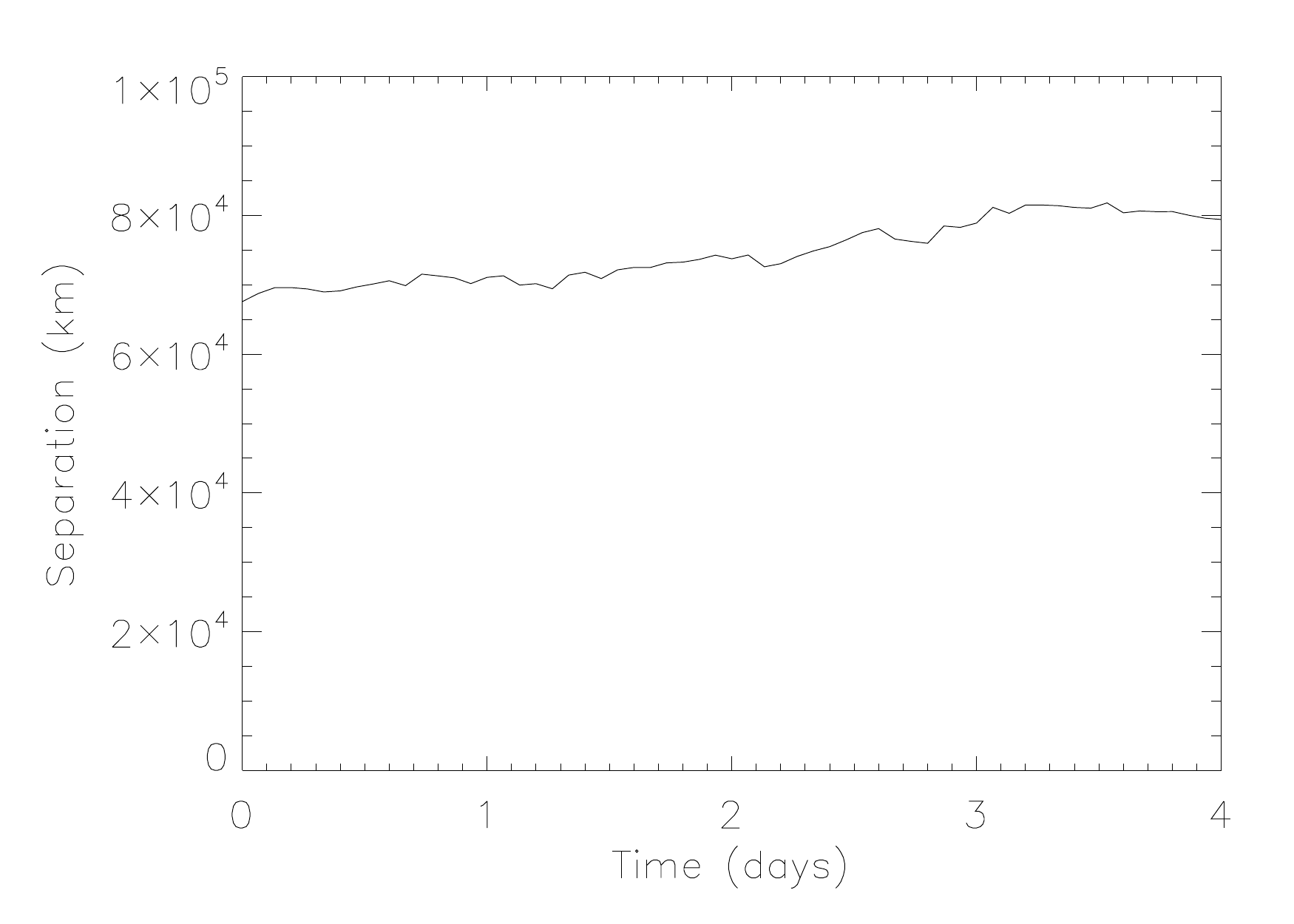}}
\caption{Separation between the center of flux of the positive and negative polarities as a
function of time in days from 16$^{th}$ December 1996 19:12 UT.}
\label{fig:fig4}
\end{figure}

On studying the magnetograms over the period of interest, the evolution of NOAA 8005 
is dominated by the dispersal of magnetic flux from the main flux concentrations. 
There is no significant emergence of new magnetic bipoles. On studying Figure~\ref{fig:fig2}
or the movie of the evolution (movie1.mpg), the motion of the magnetic elements is clearly 
dominated by a random walk due to supergranular flows and no strong shear or vortical flows are 
present. In Figure~\ref{fig:fig4} the separation distance ($S(t)$) between the centers of flux in both the
positive and negative polarities can be seen, given as a function of time ($t$). This separation is 
defined as  $S = \mid {\bf{S}}(t) \mid $ where ${\bf{S}}(t)$ is the vector pointing between the centers 
of flux in each polarity,
\begin{equation}
{\bf{S}}(t) = \frac{ \sum_{B_z > 0} B_z(i,j) {\bf{R}}_{i,j}}{\sum_{B_z > 0} B_z(i,j)} - 
\frac{ \sum_{B_z < 0} B_z(i,j) {\bf{R}}_{i,j}}{\sum_{B_z < 0} B_z(i,j)}
\end{equation}
where $B_z (i,j)$ is the line of sight component of the field at the $i^{th}$,$j^{th}$ pixel
and ${\bf{R}}_{i,j}$ is the position vector of the i$^{th}$,j$^{th}$ pixel from the origin. The
origin is 
defined to be the lower left corner of the magnetogram. Throughout the four day period, as the 
active region crosses central meridian, the separation of the polarities is seen to slightly 
increase from $68,000$km to $80,000$km. This increase is only $18\%$ that of the 
initial separation. While 
in general, the region is diverging, there are several instances where patches of strong magnetic 
flux separate from the main polarities and cancel along the Polarity Inversion Line. A key 
point to note is that during the evolution no systematic shear motions or vortical motions of the magnetic 
flux are seen. The motions of magnetic elements are mainly due to the buffeting as a result of 
supergranular motions. Although supergranular motions provide the only obvious flow 
pattern, due to the large latitudinal extent of the active region differential rotation may have a 
small non-negligible effect. The evidence and consequences of this will be discussed in Section 6.

\begin{figure}[t]
\centering\includegraphics[scale=0.5]{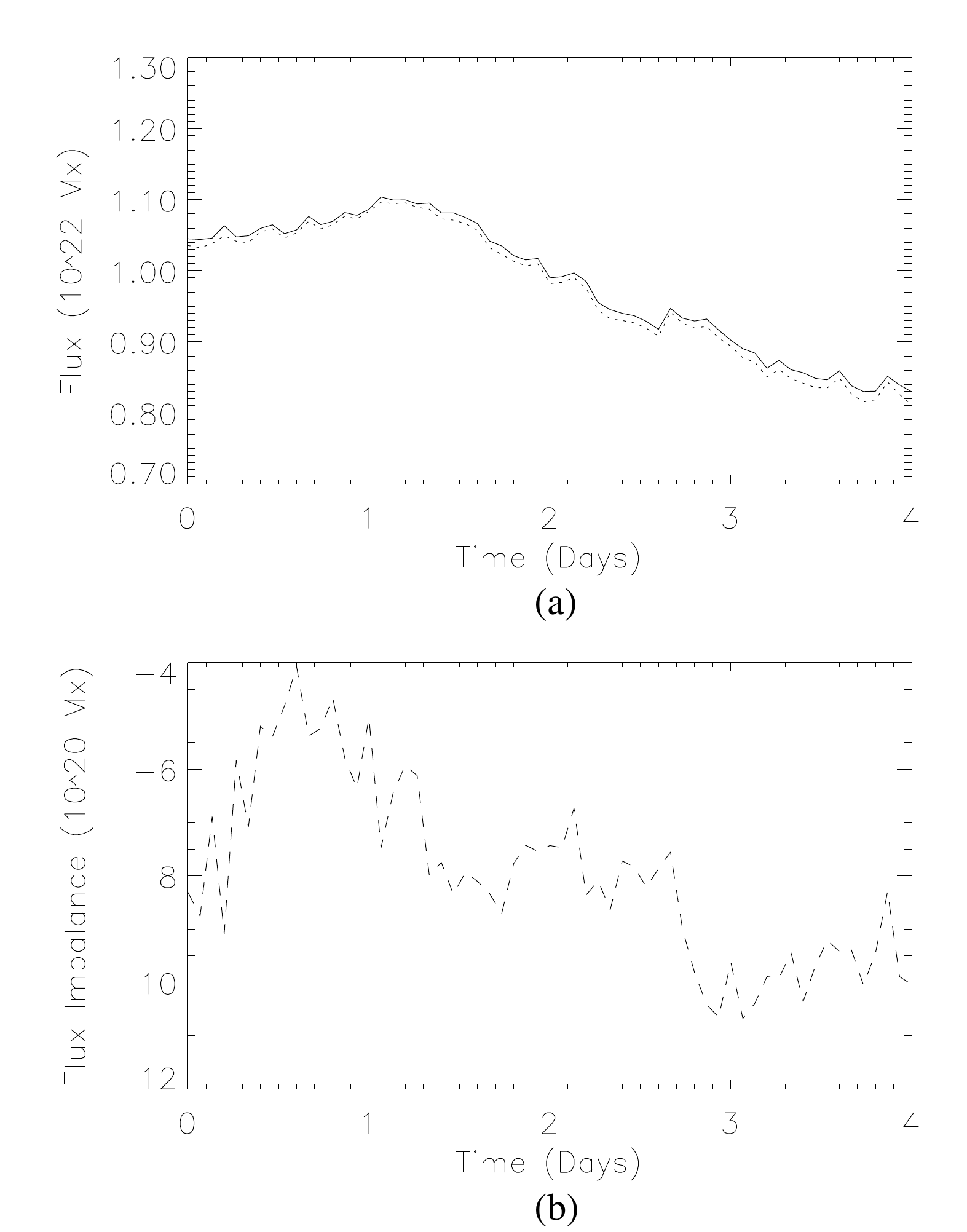}
\caption{(a) Total unsigned flux ($10^{22}$Mx) as a function of time (days). The solid line
denotes the value determined from the de-rotated data. The dashed line is once a correction
has been applied to ensure flux balance. (b) Imbalance of flux ($10^{20}$Mx) as a function of 
time (days). In both panels, time is from  16$^{th}$ December 1996 19:12 UT. }
\label{fig:fig3}
\end{figure}

In Figure~\ref{fig:fig3}(a) the variation of the total unsigned magnetic flux within the selected active 
region can be seen as a function of time. The flux values are given in units of 
$10^{22}$Mx and time in days from 16$^{th}$ December 1996 19.12.05 UT. The solid line in 
Figure~\ref{fig:fig3}(a) shows the total 
unsigned flux as derived from the individual magnetograms. The flux 
values are consistent with a medium sized active region. Initially over the first day the flux
values are seen to increase slightly, possibly due to LOS changes as the region rotates
toward central meridian and the emergence of a small bipole on the south-western edge of the active 
region. However, between days 1-4 the flux decays away as cancellation becomes significant
along the internal PIL \citep{gree09}. By the end of 
day 4 the flux has dropped by $20\%$ of its original value. This behavior fits 
well with the observation that it is a decaying active region. The peak flux densities
found within both the positive and negative polarities are around $1$kG. In 
Figure~\ref{fig:fig3}(b) the flux difference (flux imbalance) calculated for each magnetogram can be seen as a 
function of time. Over the four day period the imbalance is systematically negative, possibly due to a 
dominantly negative surrounding small scale fields. At 
worst it is $10\%$ that of the total flux of the active region. As this value is small, the active 
region may be regarded to be in flux balance to a high degree of accuracy. 

To use the magnetograms as a lower boundary condition in the numerical simulations complete flux 
balance is required. Therefore in each magnetogram the imbalance per pixel is determined and 
subtracted from each pixel. This correction varies from one magnetogram to the next. 
In general it is less than $\mid 6 \mid$G and so is comparable to the noise level within the
magnetograms
and significantly less that the peak flux density. After correction, the variation of the 
total unsigned flux is given by the dotted line in Figure~\ref{fig:fig3}(a). It can be seen that 
this line follows the uncorrected flux very closely and applying the correction does not 
significantly effect the net flux of the active region.

\section{The Model}

\subsection{Coronal Model}

To simulate the evolution of the coronal magnetic field above the active region a 
magnetofrictional relaxation technique is applied. This technique evolves the coronal magnetic 
field through a sequence of related non-linear force-free equilibria. The technique is similar 
to that described in the papers of \cite{bal00} and \cite{mac06a} 
where the code described in these papers is adapted to a local cartesian frame of 
reference. Previously the technique has been successfully applied to consider the coronal field 
based on evolving photospheric boundary conditions in either idealized setups 
\citep{mac03,mac05,mac09} or approximations to observed 
magnetograms \citep{mac00,yea07,yea08}. A key element of these simulations is 
that through considering a time sequence of evolved equilibria, this allows the memory of previous 
flux connectivities and currents to be maintained from one time to the next as energy and
magnetic helicity is injected into the corona. Such a feature is significantly different compared 
to independent extrapolations which do not maintain such memory.

\begin{figure}[t]
\centering\includegraphics[scale=0.4]{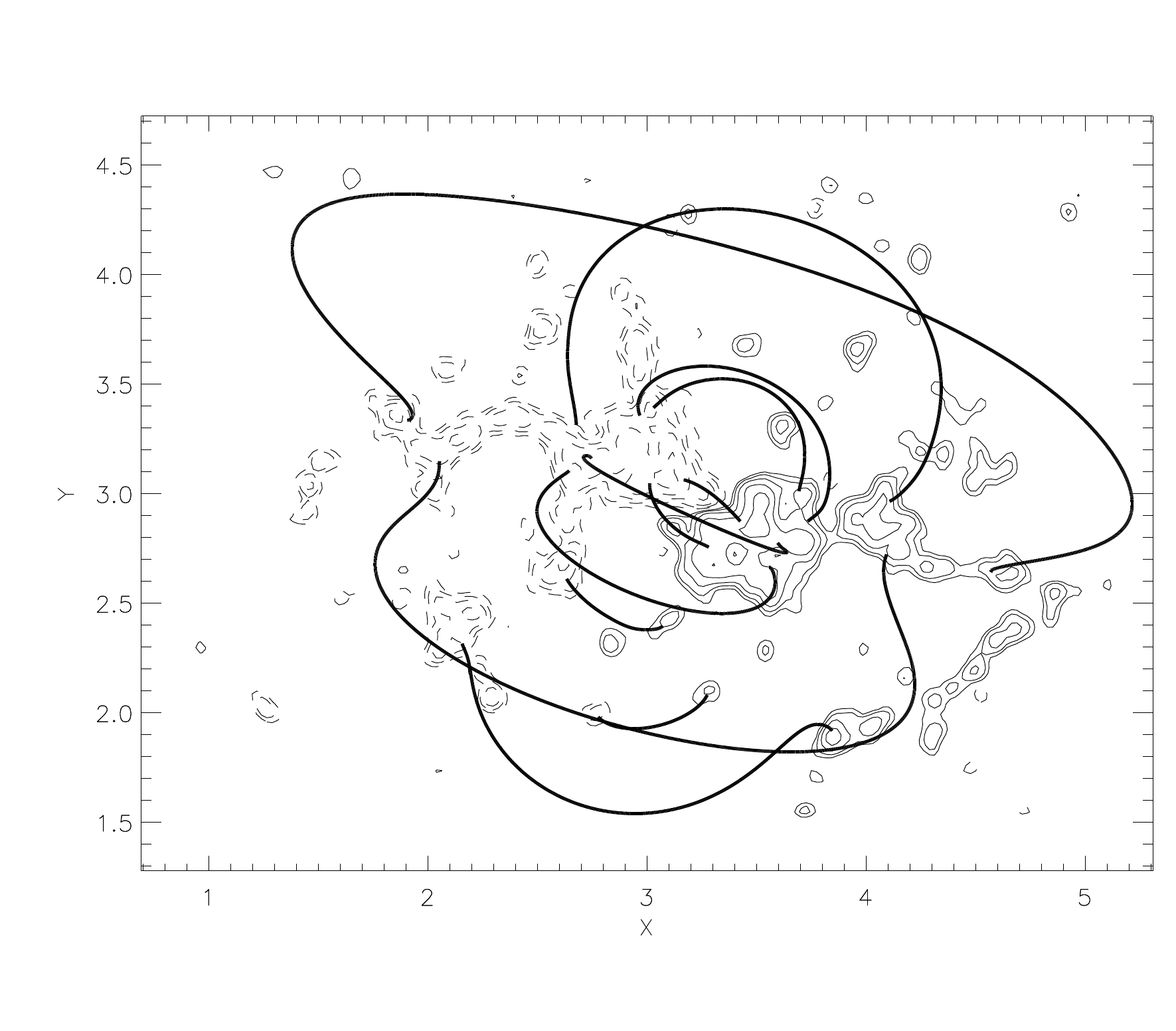}
\caption{Field line plot showing the initial potential field used in the simulation.
The thin solid/dashed lines denote positive/negative contours of the normal field component 
on the base, while the thick lines are the field lines.}
\label{fig:fig9}
\end{figure}

The magnetic field, ${\bf B} = \nabla \times {\bf A}$, is evolved by the induction equation, 
\begin{eqnarray}
\frac{ \partial {\bf{A}}}{ \partial t} = {\bf{v}} \times {\bf{B}}
\label{eq:ind}
\end{eqnarray}
where ${\bf{v}}({\bf{r}},t)$ is the plasma velocity and for the present study non-ideal terms are
not included. To ensure that the coronal field evolves through a series of force-free states a 
magneto-frictional method is employed (Yang \et, 1986). We assume that the plasma 
velocity within the interior of the box (representing the solar corona) is given by,
\begin{eqnarray}
{\bf{ v}} = \frac{1}{\nu} \frac{ {\bf{j}} \times {\bf{B}}}{ B^2}, \label{eq:lor}
\end{eqnarray}
where ${\bf{j}} = \nabla \times {\bf{B}}$ and $\nu$ is the coefficient of friction. 
Equation (\ref{eq:lor}) represents in an approximate manner, the fact that in the corona 
the Lorentz force is dominant and simulates the plasma experiencing a frictional force 
as it moves with respect to the reference frame. This ensures that as the coronal field
is perturbed via boundary motions it evolves through a series of quasi-static non-linear force-free field
distributions, satisfying ${\bf{j}} \times {\bf{B}}=0$. To carry out the 
computations a staggered grid is used to obtain second order accuracy for all 
derivatives and closed boundary conditions are applied on the side and top 
boundaries to match those given in Section 3.2. 

A key new feature within this study is that
the evolution of the coronal field is driven directly through the insertion of a time sequence
of observed line of sight component magnetograms. This means that the evolution of the
photospheric field occurs in exactly the same way as is seen in the observations. No
simplifications, idealization or smoothing is applied to the observed data and the data is 
inserted as observed (with only a small correction for flux balance applied). The technique of 
specifying the sequences of the lower boundary conditions is discussed next.

\subsection{Lower Boundary Condition and Initial Condition}

To model the evolution of the active region, 61 magnetograms taken at discrete intervals of
$96$ min are available, covering the four day period around central meridian passage. From these
magnetograms a continuous time sequence of lower boundary conditions is produced. This
sequence of lower boundary conditions is designed to match each observed magnetogram, 
pixel by pixel, every 96 minutes. However, to model the evolution of 
the coronal field, $A_{xb}$ and $A_{yb}$ the horizontal components of the vector potential {\bf A}
on the base that correspond to this magnetogram must be determined. To determine these and produce a 
continuous time sequence the following process is applied,  
\begin{enumerate}
\item Each of the observed magnetograms, $B_z(x,y,k)$ for $k=1 \rightarrow 61$ are taken, 
where $k$ represents the discrete 96 min time index.
\item Next the horizontal components of the vector potential at the base, $z=0$, are written in
the form,
\begin{eqnarray}
A_{xb} (x,y,k) = \frac{\partial \Phi} {\partial y} , \nonumber \\
A_{yb} (x,y,k) = - \frac{\partial \Phi} {\partial x} \nonumber
\end{eqnarray}
where $\Phi$ is a scalar potential.
\item For each discrete time index $k$, the equation
\begin{eqnarray}
B_{z} = \frac{ \partial A_{yb}}{\partial x} - \frac{ \partial A_{xb}}{\partial y} \nonumber
\end{eqnarray}
then becomes,
\begin{eqnarray}
\frac{ \partial^{2} \Phi}{\partial x^{2}} + \frac{ \partial^{2} \Phi}{\partial
y^{2}}= - B_{z}.
\end{eqnarray}
which is solved using a multigrid numerical method. Details of this method can be found in the 
papers by Finn \et (1994) and Longbottom (1998) and references therein. For a full description of the
boundary conditions applied see \cite{mac09}.
\end{enumerate}

\begin{figure*}[tbh]
\centering\includegraphics[scale=0.6]{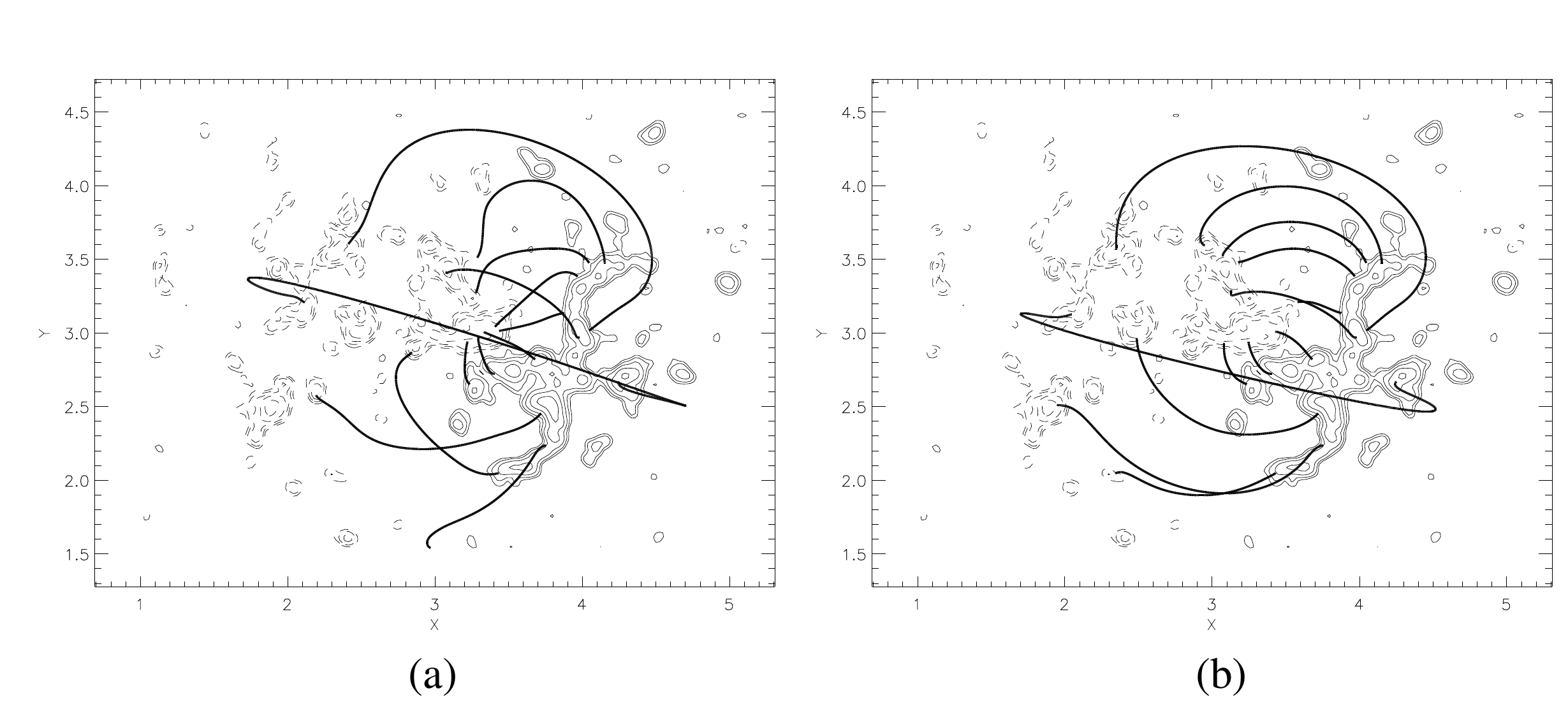}
\caption{Field line plots showing (a) the non-linear force-free field after 4 days of evolution and 
(b) a potential field corresponding to the same normal field component as that of
the non-linear force-free field after 4 days.}
\label{fig:fig10}
\end{figure*}

On solving for the scalar potential, $\Phi$ this determines the horizontal components of the
vector potential on the base ($A_{xb},A_{yb}$) for each discrete time interval, $96$ min apart. 
To produce a continuous
time sequence between each of the observed distributions a linear interpolation of
$A_{xb}$ and $A_{yb}$ between each time interval $k$ and $k+1$ is carried out. To interpolate the
fields 500 interpolation steps are used. By linearly interpolating the horizontal components
of the vector potential on the base,
this effectively evolves the magnetic field from one state to the other. Therefore techniques such as 
local correlation tracking are not required to determine the horizontal velocity.
Numerically it also means that undesirable effects such as numerical overshoot or flux pile
up at cancellation sites do not occur and no additional numerical techniques to remove these
problems have to be applied. 

The technique described above means that there are two time-scales involved in the 
evolution of the lower boundary condition. The first which is $96$ min is the time scale between 
observations, the second which is $11.52$s is the time-scale introduced to produce the advection 
of the magnetic polarities between observed states by interpolation along with the relaxation of 
the coronal field. The process described above reproduces the observed magnetograms at each 
$96$ min discrete time interval and therefore produces a highly accurate
description of the magnetogram observations. 

In movie2.mpg a comparison of the actual observations (left) and  normal magnetic field component 
produced by the above technique (right) can be seen. A
good agreement between the two is found. The new technique means that neither feature tracking, nor Local 
Correlation Tracking is required to derive horizontal velocity profiles, in order to
simulate the evolution of magnetic fields between observed distributions, therefore removing a source
of uncertainty. At this point it should be noted that the above technique only specifies
$A_{xb},A_{yb}$ on $z=0$. It however does
not specify $A_z$ which lies, $1/2$ a grid point in the box and is determined by Equation~\ref{eq:ind},
the coronal
evolution equation. Non-potential effects near the base, as a result of the evolving lower boundary
condition, may be contained within this term. These are systematically produced as the horizontal field 
components on the lower boundary evolve in time.

\begin{figure}[t]
\centering\includegraphics[scale=0.5]{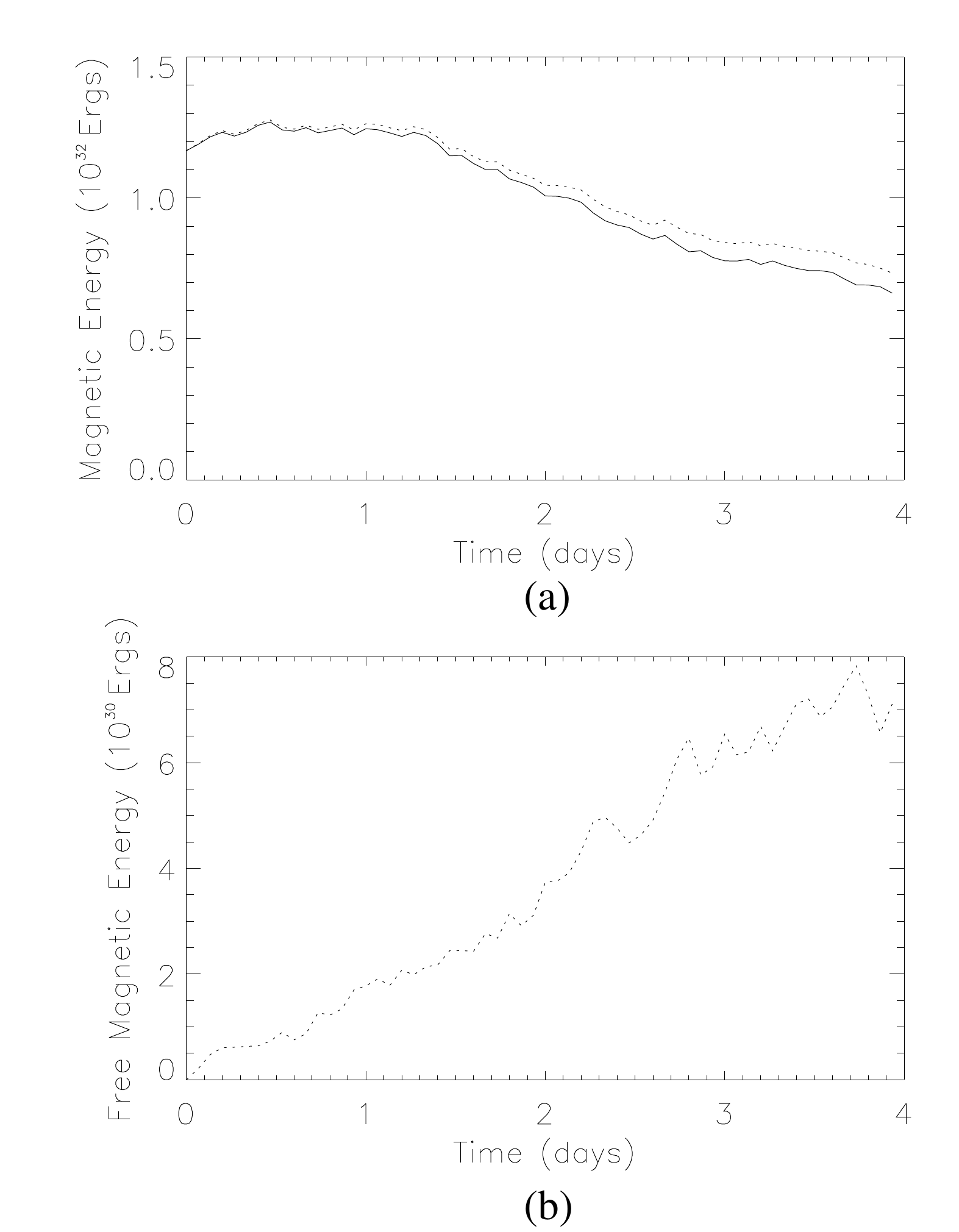}
\caption{(a) Total magnetic energy ($10^{32}$ Ergs) versus time from 16$^{th}$ December 1996 19:12 UT. 
The solid line represents the 
energy of a potential field, while the dashed line is for the non-linear force-free field. (b) Free
magnetic energy ($10^{30}$ Ergs) versus time.}
\label{fig:fig5}
\end{figure}

The choice and construction of the initial condition for the coronal field is now discussed.
For the initial condition, a wide variety of choices may be made. These range from potential to
non-linear force-free fields. One method to constrain this is to use coronal images of the active 
region to determine the field geometry. The most realistic form of the initial condition is the 
non-linear force-free field. However, since no vector magnetic field data are available to us, 
such a field cannot be constructed. On comparing various linear force-free field
solutions with coronal images we find that no single value of the force-free
parameter $\alpha$ fits the whole corona structure. Therefore as vector magnetic field data are not available, 
we choose the initial condition to be a potential field, as the solution is both unique and 
stable. By choosing this initial condition, when analysing the evolution of the active region 
we restrict ourselves to studying trends in quantities, rather than absolute values. 
In future with SDO:HMI vector field observations this restriction of only considering trends will be 
removed. In Section 6 the consequences of changing the initial condition from that of a potential
field is discussed.

To construct a potential field the equation
\begin{eqnarray}
\nabla^{2}{\bf A} = 0, \label{eq:vectff}
\end{eqnarray} 
is solved in a cube with sides ranging from $ 0 < x,y,z < 6$ on a 256$^{3}$ grid where $1$ unit = $60,733$km. 
The cube represents an isolated region of the solar surface, where flux may only enter 
or leave through the lower boundary ($z=0$). As flux may only enter/leave through the lower
boundary all field lines must start and end in the $z=0$ plane, and no field lines may leave 
through the side boundaries. This means that $B_{n}$, the normal component of ${\bf B}$, vanishes 
on each of the faces except $z=0$ (which represents the photosphere). To satisfy this it is 
assumed that the tangential components of ${\bf A}$ are zero on each of the faces, except $z=0$. 
In addition, the normal derivative of the normal component of ${\bf A}$ is set to zero on each of 
these faces. If Equation (\ref{eq:vectff}) is solved subject to the above boundary conditions it 
is straight forward to show that the solution will have $ \nabla . {\bf A} = 0$ everywhere within 
the domain (Finn \et, 1994). Thus the initial condition is a potential
field associated with the imposed normal field on the boundary with the choice of the 
Coulomb gauge ($\nabla . {\bf A} = 0$). The initial  potential field is constructed from the first
observed magnetogram at 19.12.05 UT on 16$^{th}$ December 1996.

\section{Simulation Results}

\subsection{Magnetic Field Lines}
In Figure~\ref{fig:fig9} an illustration of the field lines of the potential field used as the 
initial condition can be seen. From these field lines it is clear that the decaying active region
has a simple bipolar form. The field lines connecting between the positive flux (solid contours)
and negative flux (dashed contours) form  semi-circular loops.

In Figure~\ref{fig:fig10}(a) the illustration shows the field lines for the non-linear force-free field
after 4 days of evolution. This figure can be compared to Figure~\ref{fig:fig10}(b) where the corresponding
field lines for a potential field, deduced from the same normal field component on the base are plotted. In 
each case the starting point for the field lines is taken on the base within the positive
flux region. On comparing the two images it is clear that there are many differences in the 
connectivity and structure of the field. The main differences occur low down 
along the PIL between the two main polarities. Here the field lines of the non-potential field have a much
more sheared structure as a result of energy and helicity being injected along the field lines by
the small scale convective motions.  

Another major difference is that for the field lines lying at the southern end, the connectivity of the
field is very different from that of the potential field. This is because the connectivity
within the non-linear force-free simulation is initially defined at the start of the simulation
and is preserved throughout the simulation (except where numerical diffusion becomes large).
The largest field lines within the simulation are only slightly different as the small scale convective
motions have not been able to inject helicity along the full length of these field lines during the
time period of the simulation.

\subsection{Magnetic Energy}
In Figure~\ref{fig:fig5}(a) the graph of total magnetic energy stored within the coronal field
can be seen as a function of time. The dotted line is for the
non-linear force-free field simulation, while the solid line is for a potential field deduced
from the same normal field component on the lower boundary as that of the non-linear force-free
field (see Section 3.2). Both values are initially equal to one-another as the initial condition 
is a potential field. For both cases the total energy is of the order of $10^{32}$ ergs which 
is typical for a small active region.

Initially over the first day the energy of both the non-linear force-free field and potential field
increase. This increase is due to the increasing flux values, however after 1 day the magnetic
energy in both peaks and then starts to decrease as the level of flux decreases due to cancellation. While
both the non-linear force-free field and the potential field have a similar behaviour, it can be
seen that the non-linear force-free field has a systematically higher energy compared to the
potential field. This increased energy is due to the small scale convective motions which inject a
Poynting flux into the corona and evolve the initial coronal field away from potential.
  
In Figure~\ref{fig:fig5}(b) a graph of the total free magnetic energy within the coronal field
is plotted as a function of time. The total free energy is defined as the difference between that of
the non-potential and potential fields when integrated over the entire volume. This quantity
is always positive and by the end of 
the simulation the free magnetic energy
is approximately $8 \times 10^{30}$ ergs which is $10\%$ that of the potential field. An interesting
feature is that throughout the simulation the free magnetic energy as a result of the convective
motions increases steadily. The rate of increase is around $2.5 \times 10^{25}$ ergs s$^{-1}$ indicating that
the small scale motions deduced from the magnetograms may inject large amounts of energy into the corona. 

\begin{figure*}[t]
\centering\includegraphics[scale=0.6]{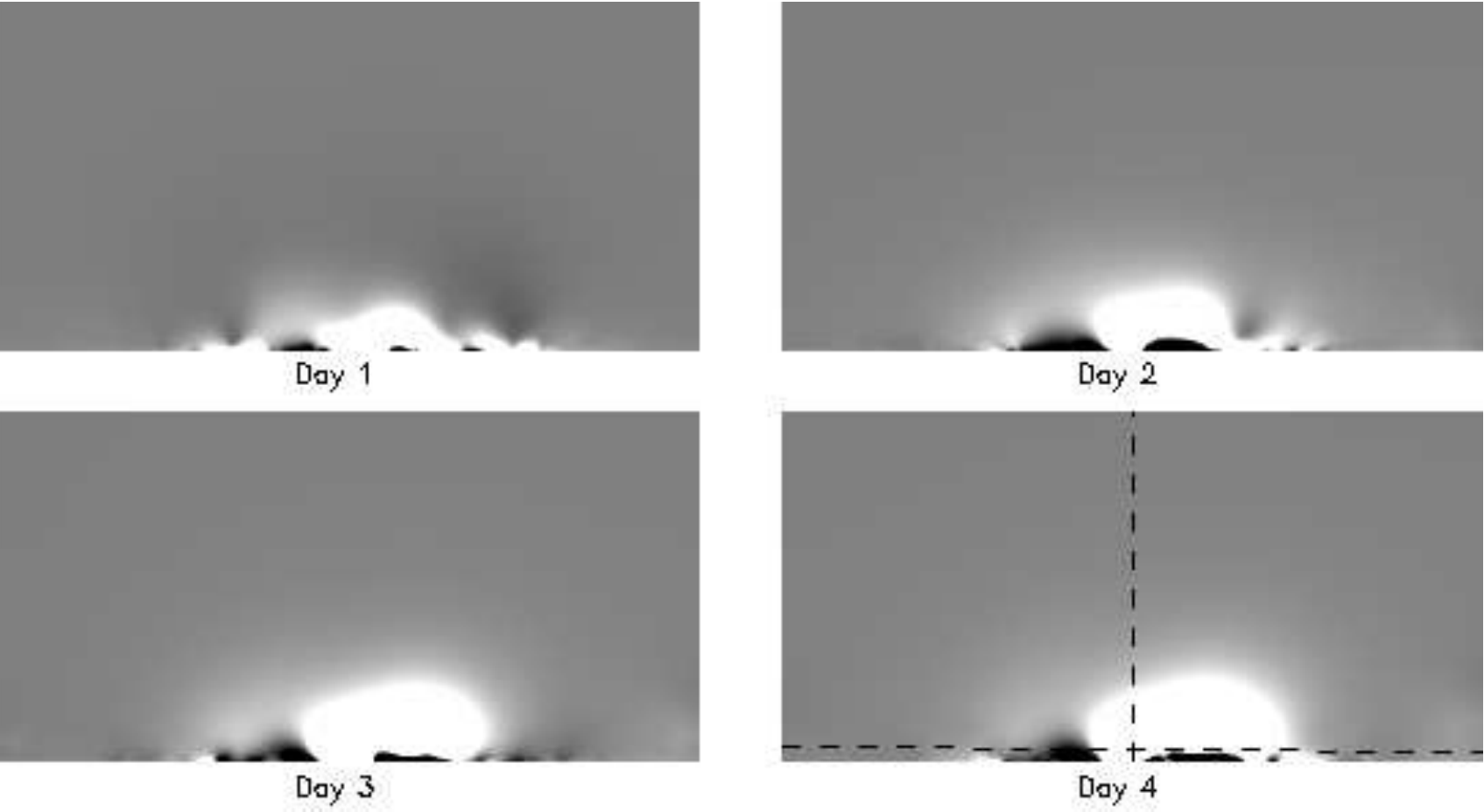}
\caption{Images showing the locations of free magnetic energy storage in the x-z plane 
summed along the y-direction for the non-linear force free field. White denotes locations where there 
is excess free magnetic energy compared to that of a potential field.  The image is set to saturate
at $\pm 1 \times 10^{27}$ ergs.}
\label{fig:fig6}
\end{figure*}

In Figure~\ref{fig:fig6} the images illustrate the locations of free magnetic energy storage
for days 1-4. The plots are in the x-z plane and give the free magnetic energy summed along 
the line of sight,
\begin{equation}
E(x,z)= A \int \frac{ ({\bf B}^2 - {\bf B_p}^2)}{8 \pi} dy
\label{eq:energy}
\end{equation}
where ${\bf{B}}$ is the magnetic field of the non-linear force-free field and ${\bf{B_p}}$ the magnetic field
of the potential field satisfying the same normal field components on the boundaries. The factor $A$
represents the area of the column being summed over ($A = 2.02 \times 10^{16}$ cm$^2$). The area factor is 
included so that the free energy along the line of sight is computed in units of ergs. In each plot
the x-direction represents the full length of the computational box, but the z-direction is only half the 
height. White locations denote where the non-linear force-free field has a higher energy than 
the potential field when integrated along the line of sight (i.e locations where 
free magnetic energy is stored). Black denotes where the non-linear force-free field has a lower energy
when integrated along the line of sight (i.e locations where there is no
free energy). We note that while the non-linear force-free field must and does have a volume integrated 
energy that is greater than that of the potential field (see Figure~\ref{fig:fig5}), there is no restriction 
that in any sub-volume that this must always be true. Hence the negative values of the line integral
quantity in Equation~\ref{eq:energy} are physically valid.

\begin{figure*}
\rotatebox{270}{\centering\includegraphics[scale=0.7]{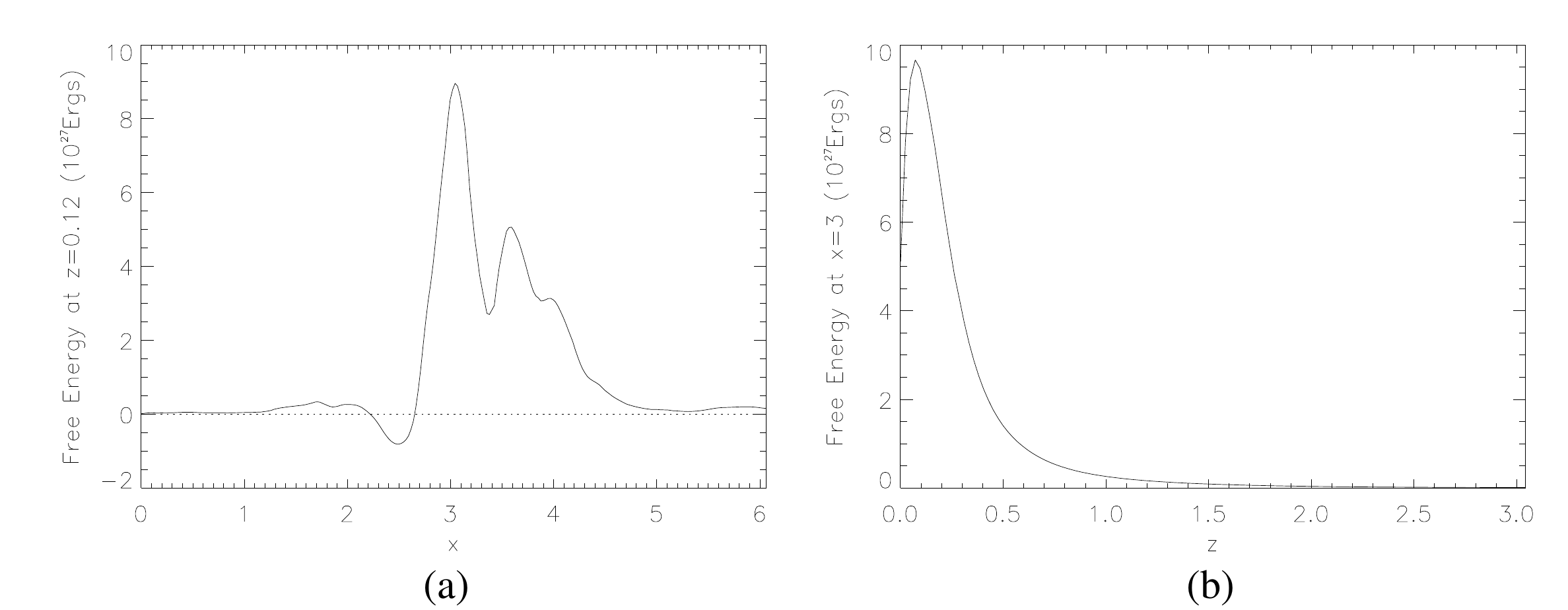}}
\caption{Graphs showing the free magnetic energy in units of $10^{27}$ ergs as a function of (a) $x$ at
$z=0.12$ units and (b) $z$ at $x= 3$ units along the dashed lines shown in Figure~\ref{fig:fig6}. Positive 
values denote locations where the non-linear force-free field energy has a higher energy compared to that 
of a potential field.}
\label{fig:fig12}
\end{figure*}

As the small scale convective motions advect the field between days 1 and 4, the locations of free magnetic 
energy storage expand up into the coronal volume and mainly lie in the center of the box.  From these images 
it is clear that the main locations of free magnetic energy is low down in the corona (below $30$ Mm) and 
between the two main flux concentrations.  It is built up at this location, as here the magnetic 
fields are the strongest and the energy injection through Poynting Flux the greatest. In addition,
these field lines are the shortest and occupy a smaller volume. 

While the images shown in Figure~\ref{fig:fig6}
are useful for illustrating the general spatial locations of free magnetic energy. Often such scaled images
lead to misleading interpretation on the exact values. To quantify this, in Figure~\ref{fig:fig12}, graphs of the
variation of free magnetic energy can be seen for day 4, along the paths given by the dashed lines in 
Figure~\ref{fig:fig6}. In Figure~\ref{fig:fig12}(a) the free magnetic energy can be seen as a function of
x at a height of $7,300$ km ($z=0.12$ units). Positive values denote locations where the non-linear 
force-free field has a higher energy, negative values where it is lower (when integrated along the line
of sight). From this graph it is clear that in the 
coronal volume the free energy is mainly positive. Near the center of the domain there is a small region where 
the non-linear force-free field has a lower energy than the potential field. However the negative value 
of the line of sight integrated free energy is small 
compared to the locations where it is positive. From this it can be seen that the scaling and color shown in  
Figure~\ref{fig:fig6} exaggerates the level where the free magnetic energy is negative. 
In Figure~\ref{fig:fig12}(b) the variation of free magnetic energy can be seen as a function of $z$,
taken at $x=3$ units. It is clear that along this cut the free energy is always positive. In general
the amount of free energy available decreases with height in the box.

\begin{figure}[t]
\centering\includegraphics[scale=0.6]{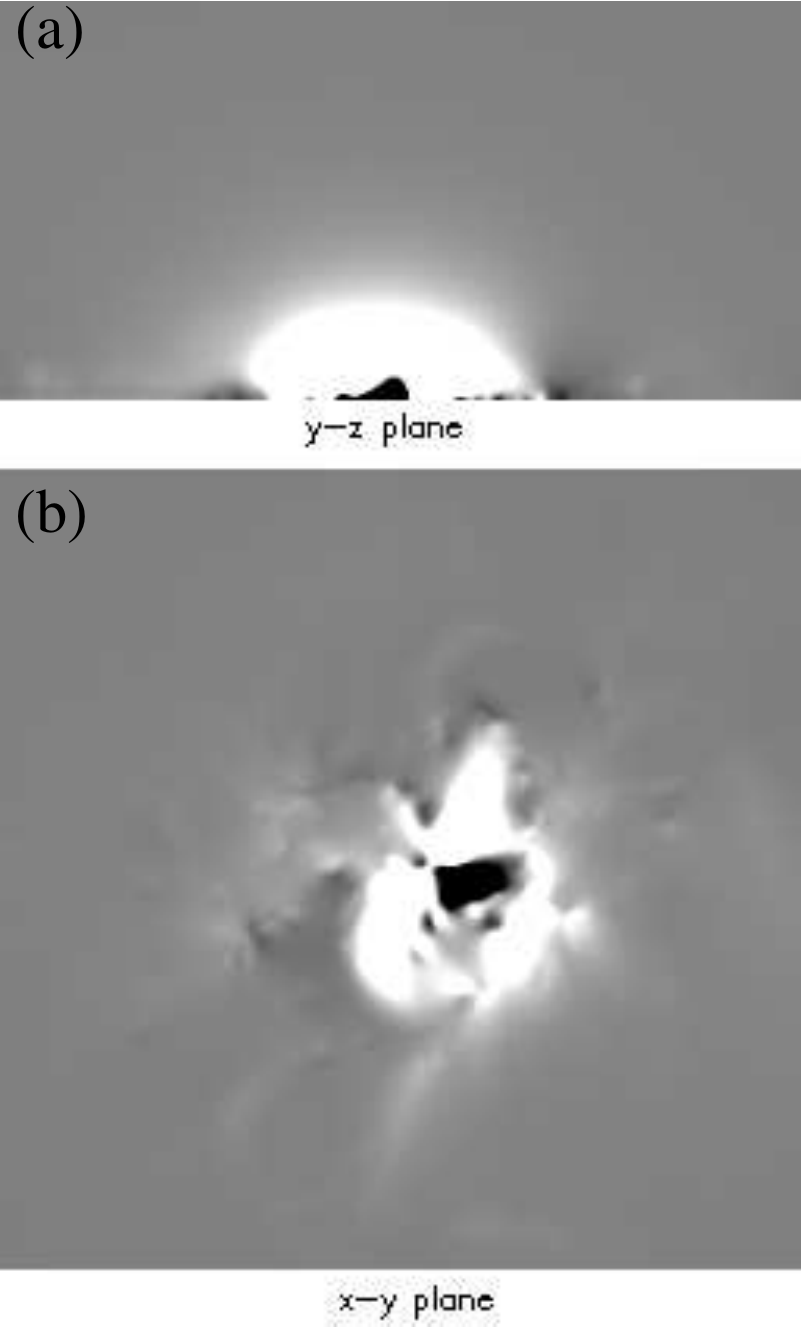}
\caption{Images showing the locations of free magnetic energy storage on day 4 in (a) the $y-z$ plane 
summed along the $x$-direction and (b) the $x-y$ plane summed along the $z$-direction for the non-linear 
force free field. White denotes locations where there is excess free magnetic energy compared to that of 
a potential field. The same scaling is used as in Figure~\ref{fig:fig6}.}
\label{fig:fig11}
\end{figure}

Alternative views of the free magnetic energy storage can be seen in Figure~\ref{fig:fig11}. 
In Figure~\ref{fig:fig11}(a) it is shown in the $y-z$ plane summed along the x-direction 
(Equation~\ref{eq:energyx}) and in Figure~\ref{fig:fig11}(b) for the $x-y$ plane 
summed along the z-direction (Equation~\ref{eq:energyz}).
In each case results are shown for day 4 and the same scaling is used as in Figure~\ref{fig:fig6}.
\begin{eqnarray}
E(y,z) = A \int \frac{ ({\bf B}^2 - {\bf B_p}^2)}{8 \pi} dx  \label{eq:energyx}\\
E(x,y) = A \int \frac{ ({\bf B}^2 - {\bf B_p}^2)}{8 \pi} dz \label{eq:energyz}
\end{eqnarray}
 
\begin{figure}[t]
\centering\includegraphics[scale=0.45]{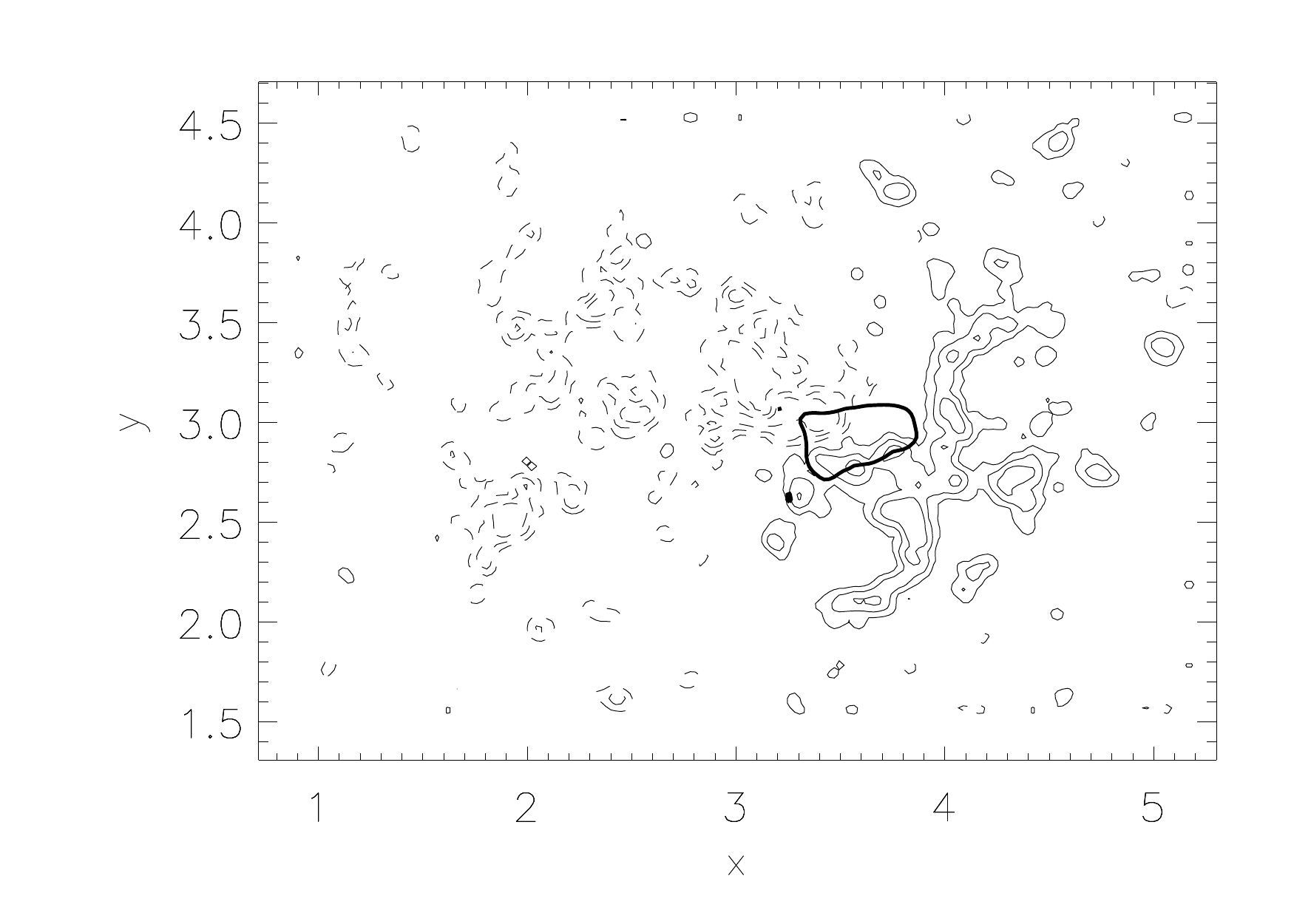}
\caption{Comparison of the underlying magnetic field distribution with locations were the non-linear force-free
field has less energy than the potential field on day 4. In the plot positive and negative photospheric magnetic flux
is given by the thin solid and dashed contours. The thick solid contour denotes the area where
the non-linear force-free field has an energy less that $6 \times 10^{26}$ergs of the corresponding potential field
when summed in the z-direction.}
\label{fig:fig13}
\end{figure}

Once again it is clear that the locations of free magnetic energy storage exist in the center of the 
computational box where the main flux concentrations are at their strongest. In 
Figure~\ref{fig:fig11}(b) the main spatial location where the line of sight integrated free
energy is negative is surrounded by positive free energy locations. 
To determine why such a region exits, Figure~\ref{fig:fig13} 
compares the photospheric magnetic field distribution of the active region to the location where the line of
sight
free magnetic energy is negative. In this plot thin solid lines denote positive flux, dashed lines 
negative flux and the thick solid line where the line of sight integrated free energy 
is $-6 \times 10^{26}$ ergs or less. It can be clearly seen that the negative region
is spatially correlated with the cancellation site between the positive and negative polarities. Therefore it 
is due to the removal of photospheric and coronal flux at this location as a result of flux cancellation
(Figure~\ref{fig:fig3}).

\subsection{Magnetic Helicity}
As the coronal magnetic field is subjected to the small scale random motions, these motions not 
only inject magnetic energy into the field but also magnetic helicity \citep{dem09}. This is seen by the
increasing complexity of the field. To quantify this injection a calculation
of relative magnetic helicity ($H_r$) is carried out. This quantity is defined as

\begin{equation}
H_r = \int_v {\bf{A.B}} d\tau - \int_v {\bf{A_p.B_p}}d\tau .
\end{equation}
where $\bf A $ and $ \bf B $ are the vector potential and magnetic field of the non-linear force-free field.
Similarly $\bf A_p$ and $\bf B_p$ are the vector potential and magnetic field of a potential field which
satisfies the same normal field components as that of the non-linear force-free field on all boundaries 
(namely $B_z$ on $z=0$ and $B_n = 0$ on all other faces). The definition of relative magnetic helicity
was first introduced by \cite{ber84} as an invariant and therefore meaningful measure of 
magnetic helicity. It should be noted that in constructing the potential field for the   
relative helicity calculation the technique described in Section 3.2 is 
used. The horizontal components of the vector potential on the base, $A_{xb}$ and $A_{yb}$, are    
identical for both the non-linear force-free and potential magnetic
fields. This means that the expression for the relative helicity does
not include an addition surface integral term.

\begin{figure}[t]
\rotatebox{270}{\centering\includegraphics[scale=0.4]{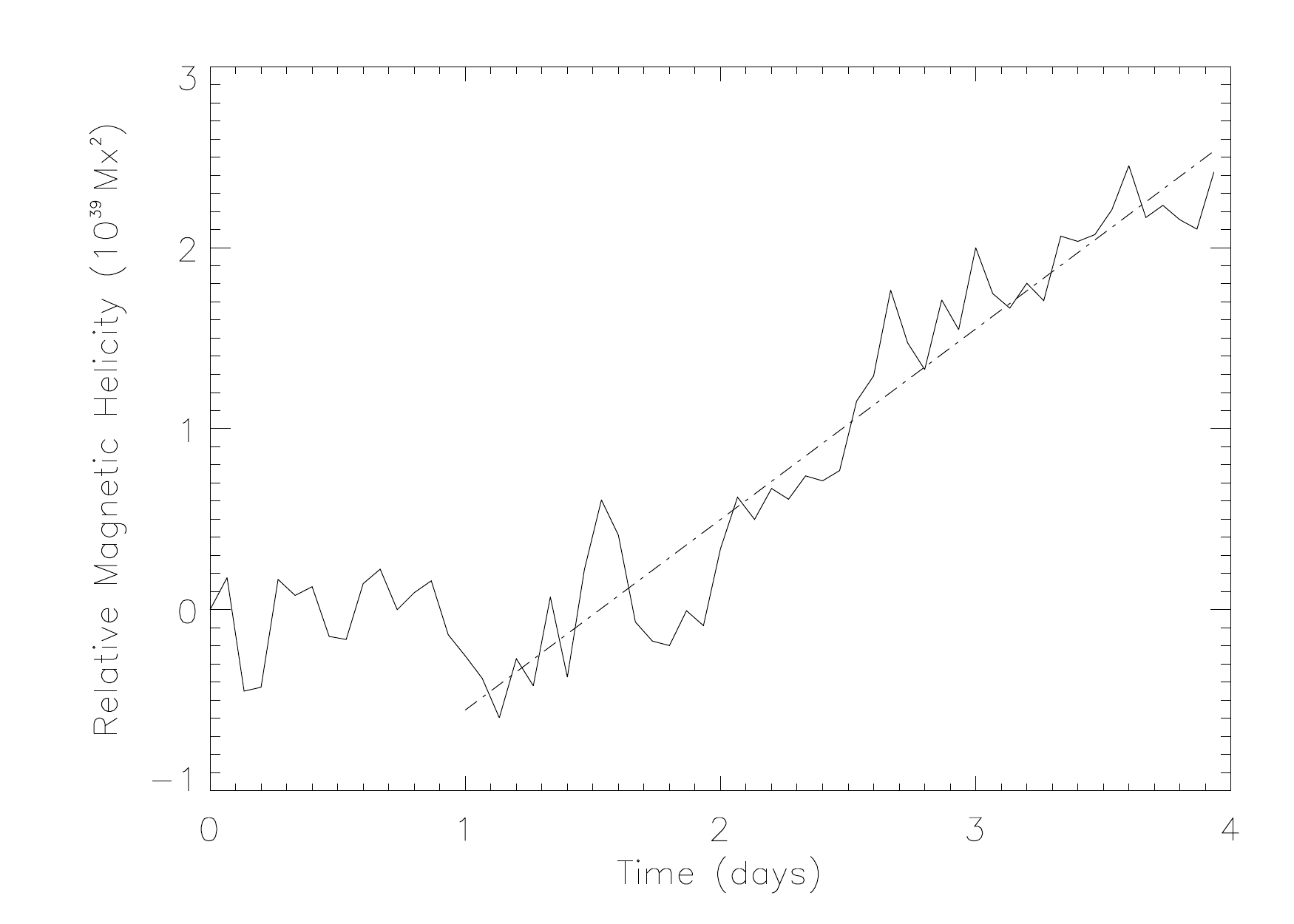}}
\caption{Variation of relative magnetic helicity within the simulation as a function of time (solid
line). The dash-dot
line denotes a straight line fit to the positive trend seen between days 1-4. The gradient of the straight 
line gives an average helicity injection rate over days 1-4 of $1.218 \times 10^{34}$ Mx$^2$s$^{-1}$.}
\label{fig:fig7}
\end{figure}

In Figure~\ref{fig:fig7} the graph shows the variation of the relative magnetic helicity as a function
of time (solid line). From the behaviour of the curve it can be seen that over the first day there is no
definite trend of helicity injection by the surface motions. The relative helicity oscillates
between positive and negative values. In contrast, between days 1-4 there is a definite trend of positive 
helicity injection, with the relative helicity showing an increasing positive value. Positive helicity
injection is consistent with observations that show that active regions in the southern
hemisphere have a dominant positive helicity \citep{pevtsov95}. The dash-dot
line in Figure~\ref{fig:fig7} is a linear best fit to the relative helicity curve between days 1-4.
The gradient of the line gives a helicity injection rate of $1.218 \times 10^{34}$ Mx$^2$s$^{-1}$. Such
an increase of helicity within the coronal field can only be as the result of helicity injection through
the lower boundary as a result of the relative motion of the magnetic fragments \citep{dem03,dem09}
as during the period of observations there is no increase in magnetic flux.  
However, through 
considering the evolution of the magnetic elements in the magnetogram, there is no clear indication as to 
the origin of the helicity injection, as the dominant motion acting on the active region are small scale 
random motions. 

In contrast, on comparing Figure~\ref{fig:fig7} to the graphs of flux variation (Figure~\ref{fig:fig3})
and {\bf{separation}} between the centers of flux elements (Figure~\ref{fig:fig4}), it can be seen that
the positive helicity injection may possibly be related to large-scale properties of the evolution
of the active region. To begin with on day 1 the flux peaks, after which it systematically 
starts to decrease. This decrease is the result of first convergence and then cancellation of the 
polarities along the internal PIL of the bipole. Secondly, around the same time the {\bf{separation}}  of the 
sources increases, so there is an overall divergence 
of the active region polarities. 
Either of these two processes (convergence/divergence) may result in the net positive helicity injection. 
However, within the present simulation we are unable to determine which. To compute this techniques such 
as Local Correlation Tracking would have to be applied
\citep{chae01,pariat05,jeong07}. While there is no clear indication of its origin, a third possibility
is that such a positive 
injection of helicity is consistent with the effect of differential rotation as will be discussed in 
Section 6.

While the relative helicity has a positive value the amount of helicity is small. Active
regions containing fluxes of $1 \times 10^{22}$Mx may well contain helicity on the
order of $10^{42} - 10^{44}$Mx$^2$ \citep{tian08}. On studying the coronal field in more detail 
it is found that the random motions inject large amounts of both positive and negative helicity, but in 
near equal amounts over the four days of the simulation. This is what is expected from such motions, therefore
resulting in a low net helicity.

\begin{figure*}[t]
\centering\includegraphics[scale=1.0]{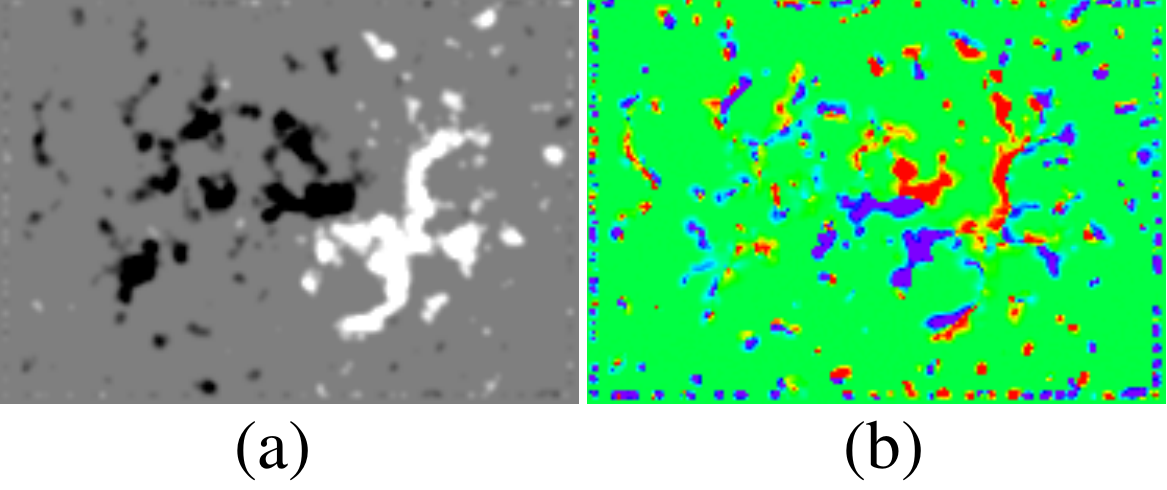}
\caption{Comparison of (a) magnetic flux distribution and (b) distribution of the force-free parameter 
$\alpha$ on the base ($z=0$) at the end of day four. The magnetic field values are set to saturate at
$\pm 100$G while the values of alpha saturate at $\pm 2.73 \times 10^{-9}$ $m^{-1}$.}
\label{fig:fig8}
\end{figure*}

In Figure~\ref{fig:fig8} a comparison of (a) the normal magnetic field component on the lower 
boundary and (b) the distribution of $\alpha$ on the lower boundary can be seen for the final time snapshot
of the simulation. Similar plots were found at other times. In 
Figure~\ref{fig:fig8}(a) white/black represents positive/negative flux, while in Figure~\ref{fig:fig8}(b)
red/blue denotes positive/negative values of $\alpha$. The magnetic field values are set to saturate
at $\pm 100$G while the $\alpha$ values saturate at $\pm 2.73 \times 10^{-9}$ m$^{-1}$. On comparing the
distributions it can be seen that the spatial distribution of $\alpha$ is well correlated to that of
the magnetic field.  The distribution of $\alpha$ take two forms. First, in locations of strong field
concentrations the $\alpha$ distribution takes the form of large coherent patterns. Secondly, it takes
the form of small isolated patches of intermingled sign which lie in small flux concentrations. It is clear
from comparing the two images, that within the large concentrations of flux, a single magnetic element
of one polarity may contain both positive and negative values of $\alpha$. Such a mixture of positive and
negative values of $\alpha$, as a result of small scale convective motions, support the results
that the  coronal field of the AR cannot be modeled by a linear force-free field as found in Section 2.2.
The positive and negative
$\alpha$ distributions are similar to the divided distribution of current used in the idealised model of
\citet{reg09}. An interesting feature of the
comparison is that within the large flux concentrations the northern portions appear to contain
more positive values of $\alpha$, while the southern portions have more negative values. In addition to the
$\alpha$ values that lie within the center of the simulation, values may be seen along the edge of the
magnetogram area. Such edge values are a pure boundary effect and should not be considered
further.

For the active region considered here no vector magnetogram data are available in order to compare
observed $\alpha$ distributions or trends of $\alpha$ in the observations with those obtained within the 
simulation. In future with the launch
of SDO and the SDO:HMI instrument, regular vector magnetic field data will be available which can be compared
to future simulations applying this technique. Through this we will be able to determine how much 
helicity is injected due to small scale motions, and in addition how much must be injected
by other effects (e.g. torsional Alfven waves). This provides a new tool for the quantification
of such effects which may be applied directly to observations.

\section{Simple Calculations}
To determine if the rate of energy build-up within the present simulation is realistic, a simple order 
of magnitude calculation is carried out 
to determine the rate of inflow of energy into the lower boundary via the Poynting Vector, 
${\bf S} = (c/4 \pi) {\bf E} \times {\bf B}$. The Poynting flux ($P$) through the lower boundary is,
\begin{equation}
P = - \frac{c}{4 \pi} \int_s ({\bf E} \times {\bf B}).{\bf{da}}
\end{equation}
where ${\bf E}$ is the electric field and ${\bf{da}}= da {\hat{\bf n}}$. In order of magnitude terms
this equation may be simplified to,
\begin{equation}
P \approx \frac{v_h B_z B_h A_m}{4 \pi}
\end{equation}
where $v_h$ is the typical horizontal velocity, $B_z$ a typical normal field component, $B_h$ a typical
horizontal field strength and $A_m$ the area of energy input. The area of input is taken as the area of the
observed magnetogram $A_m = 4.7 \times 10^{20}$ cm$^2$, $v_h = 0.5 kms^{-1}$ the peak flow rate of a
supergranular cell. The values of the average field components are then taken to be $B_z = 12$G and $B_h = 2$G, 
where these values are determined from the initial potential field. Through this an upper estimate of the 
Poynting flux becomes $4.5 \times 10^{25}$ ergs s$^{-1}$. This
value is around a factor of 2 greater than the value deduced within the simulations of 
$2.5 \times 10^{25}$ ergs s$^{-1}$. Although the order of magnitude calculation is larger by a factor of 2
this value is an over estimate as it assumes  (i) the flow is always at the peak rate and
(2) flows that are steady and systematic. In reality all elements in the magnetogram do not move at all times 
and in the simulations the flow is irregular. Taking this into account it can
be seen that the simulation produces a realistic energy input into the coronal volume. 

In the second calculation, we consider how important the energy input from the small convective motions
is relative to an estimate of the radiative losses for the active region. The radiative losses
in an optically thin plasma at coronal temperatures is given by,
\begin{equation}
E_R = n_e^2 R(T) ~~ \mathrm{ergs cm^{-3} s^{-1}}
\end{equation}
where $n_e$ is the electron number density, $T$ the temperature and $R(T)$ the radiative loss function 
which is given by Rosner et al (1978) as,
\begin{equation}
R(T) \approx 10^{-17.73} T^{-2/3}, ~~ \mathrm{6.3 < log~T < 7.0}
\end{equation}
The average temperature of the active region is determined to be $log~ T = 6.41$ through using the SXT filter
ratio technique on the Al.1 and AlMg filters. We take the volume of the coronal emitting plasma to be
$10^{29}~$cm$^3$. Note that this is not the volume of the computation box but rather an upper estimate of
the volume of emitting plasma in the active region. Taking also an upper estimate of the electron
number density to be $10^9$cm$^{-3}$, provides an upper estimate of the coronal radiative losses as 
$\sim 1 \times 10^{25}$ ergs s$^{-1}$. This value can be seen to be of the same order as the energy build
up due to the small scale convective motions. 

While the two values are consistent with one-another, it should be noted that the free magnetic energy in the 
simulation is held within the coronal field and therefore not immediately accessible to heat the corona and 
account for the radiative losses. For such a process to take place an additional energy release mechanism, 
would have to act. Possible mechanisms include nanoflares \citep{Parker88,Browning08,Wilmot-Smith10} or
turbulent reconnection\citep{heyv84}. At 
this point it is unclear if such a mechanism may release the energy. Also the simple calculation does not 
take into
account the effect of thermal conduction which is an important factor in the energy balance of hot coronal loops. 
However, taking these limitations into account, the calculation does show that convective motions may play a key 
role in the energy build up of the corona.

Using the rates of energy injection determined in the simulation and assuming that it remains 
constant over the 
life-time of the active region, it is possible to estimate how much free energy the active region could 
have on the 16th December, the start day of the present simulations, due to prior evolution as it rotated 
onto the disk and towards central meridian. Assuming that it emerged just before
disk passage the active region would have evolved for 5 days. As a result the small scale convective motions 
would inject a total of $1.1 \times 10^{31}$ergs of free magnetic energy into the coronal field.
This value is 10$\%$ of the active region energy on the 16th December. While the energy 
may have been $10\%$ higher, we cannot use this to constrain the initial condition without having vector magnetic 
field data to constrain the distribution of $\alpha$ on the lower boundary.

\section{Discussion and Conclusions}
In this paper a new technique for modeling non-linear force-free fields directly from line of sight
component magnetogram observations has been presented. The key new feature of this technique is that sequences of 
line of sight magnetograms can be directly used as lower boundary conditions to drive the evolution of 
coronal magnetic fields between successive force-free equilibria. As a result, the lower boundary
condition in the model closely follows the observed magnetic field evolution as 
given by the observed line-of-sight magnetograms over the period of observations.

The technique is illustrated by applying it to MDI observations of a decaying active region, NOAA  AR 8005. The 
active region was close to central meridian and in flux balance to a high degree. It was modeled over a 4
day period, where 61 individual SoHO:MDI 96 min magnetograms where available to use as lower boundary 
conditions. During this time the dispersal of the active region was mainly dominated by random motions due 
to small scale convective cells and flux cancellation along the internal polarity inversion line. 

Through applying the 61 magnetograms as lower boundary conditions and simulating a continuous evolution
of the surface and coronal field between them, the build up of free magnetic energy into the corona was studied. 
It should be noted here that the initial condition for the coronal field in the simulation was taken to
be a potential field. Due to the availability of only line-of-sight component magnetograms this was the only
unique choice. However at the time corresponding to the start of the simulation the actual coronal
field of the active region need not have been potential. Therefore our 
aim was not to reproduce the evolution of the coronal loops as seen in coronal observations, but rather to 
study the trends of energy input over the period of four days, where this is deduced from using observed 
magnetograms.

When the initial coronal field is taken to be potential, the small-scale 
motions inject around $2.5 \times 10^{25}$ ergs s$^{-1}$ of free magnetic energy, this energy is mainly 
stored in the low corona below 30 Mm. However, even if the initial state on the Sun was different,
this would not significantly alter the trends of energy injection found within the simulation. 
To illustrate this
we have repeated the simulation using an initial condition of a linear force-free field, with $\alpha$
values of $1.55 \times 10^{-8}$m$^{-1}$ and $1.86 \times 10^{-8}$m$^{-1}$. These values are 80$\%$ and
88$\%$  of the first resonant value of $\alpha$. Only positive values of $\alpha$ are chosen as the
active region lies in the southern hemisphere. With these alternative initial conditions similar results
and trends are found to those shown in Figure~\ref{fig:fig5}. The only difference is that the magnetic
field has a higher initial energy. The subsequent rates of energy injection are
$2.65 \times 10^{25}$ ergs s$^{-1}$ and $2.8 \times 10^{25}$ ergs s$^{-1}$ respectively. These values
are consistent to those found when using the potential field. The slightly higher values occur due to the 
increased horizontal component of field in the linear-force free field leading to a higher Poynting flux. 
For all cases the energy
injection through this new technique is consistent with estimates derived from poynting flux calculations. After
just 4 days of evolution the free energy is 10$\%$ that of the potential field. Such energy buildup, is 
sufficient to explain the radiative losses at coronal temperatures within the active region
($1 \times 10^{25}$ ergs s$^{-1}$). Therefore small 
scale convective motions play an integral part in the energy balance of the corona.

Through studying the evolution of the relative magnetic helicity in the coronal field, it was found that near 
equal amounts of positive and negative helicity are injected into the coronal field, but with a slight preference
for positive helicity injection in the later stages. This is consistent with the active region being buffeted
by small scale random motions and observations that show that active regions in the southern hemisphere
have a preference for positive helicity \citep{pevtsov95}. One possible cause of this slight positive 
helicity 
injection is the effect of differential rotation. In the southern/northern hemisphere differential rotation
will inject positive/negative helicity into the corona \citep{DeVore2000}. \cite{dem02} showed that helicity
injection by differential rotation is composed of two terms, which for a decayed and dispersed active region
are of opposite sign. The injection rate also increases with active region
latitude and varies with active region tilt angle. During the short time period of our study, where the
active region lies near the equator, we expect differential rotation to play only a minor role in
ejecting positive helicity into the active region. 

The new technique presented here has wide ranging applications for modeling the evolution of photospheric 
magnetic fields observed on the Sun and the subsequent effect these have on the coronal field over long 
periods of time. In future we shall use new high resolution, high cadence full disk vector magnetic field data
from SDO:HMI in combination with coronal observations to  constrain the initial 
condition. With this we may then compare the evolution of the coronal magnetic field
with that seen in coronal observations. Under such constraints it will 
become a powerful tool to study the nature and properties of coronal magnetic fields. \\

DHM would like to thank UK STFC for their financial support and the Royal Society for funding 
a 81 processor supercomputer under their Research Grants Scheme.  LMG would like to thank
the Royal Society  and Leverhulme Trust for financial support. We acknowledge the use of data provided by 
the SoHO/MDI instrument. The authors would like to thank the referee for his constructive comments which 
have improved this paper.


\begin{thebibliography}{}
\bibitem[Archontis et al.(2004)]{arch04} 
Archontis, V., Moreno-Insertis, F., Galsgaard, K., Hood, A., \& O'Shea, E.\ 2004, \aap, 426, 1047 

\bibitem[Benz(2008)]{benz08} 
Benz, A.~O.\ 2008, Living Reviews in Solar Physics, 5, 1 

\bibitem[Berger \& Field(1984)]{ber84} 
Berger, M.~A., \& Field, G.~B.\ 1984, Journal of Fluid Mechanics, 147, 133 

\bibitem[Bobra et al.(2008)]{bob08} 
Bobra, M.~G., van Ballegooijen, A.~A., \& DeLuca, E.~E.\ 2008, \apj, 672, 1209 

\bibitem[Browning et al.(2008)]{Browning08} 
Browning, P.~K., Gerrard, C., Hood, A.~W., Kevis, R., \& van der Linden, R.~A.~M.\ 2008, \aap, 485, 837 

\bibitem[Chae(2001)]{chae01} Chae, J.\ 2001, \apjl, 560, L95 

\bibitem[Charbonneau(2005)]{charb05} 
Charbonneau, P.\ 2005, Living Reviews in Solar Physics, 2, 2 

\bibitem[Cremades et al.(2006)]{crem06} 
Cremades, H., Bothmer, V., \& Tripathi, D.\ 2006, Advances in Space Research, 38, 461 

\bibitem[D{\'e}moulin et. al(2002)]{dem02} 
D{\'e}moulin, P., Mandrini, C.~H., van Driel-Gesztelyi, L., L{\'o}pez Fuentes, M.~C. \& Aulanier, G.\ 2002, \solphys, 207, 87 

\bibitem[D{\'e}moulin \& Berger(2003)]{dem03} 
D{\'e}moulin, P., \& Berger, M.~A.\ 2003, \solphys, 215, 203 

\bibitem[D{\'e}moulin \& Pariat(2009)]{dem09} 
D{\'e}moulin, P., \& Pariat, E.\ 2009, Advances in Space Research, 43, 1013 

\bibitem[De Rosa et al.(2009)]{derosa09} 
De Rosa, M.~L., et al. 2009, \apj, 696, 1780

\bibitem[DeVore(2000)]{DeVore2000} DeVore, C.~R.\ 2000, \apj, 539, 
944 

\bibitem[Fan(2009)]{fan09} 
Fan, Y.\ 2009, \apj, 697, 1529 

\bibitem[Galsgaard et al.(2007)]{gals07} 
Galsgaard, K., Archontis, V., Moreno-Insertis, F., \& Hood, A.~W.\ 2007, \apj, 666, 516 

\bibitem[Green \& Kliem(2009)]{gree09} 
Green, L.~M., \& Kliem, B.\ 2009, \apjl, 700, L87

\bibitem[Heyvaerts \& Priest(1984)]{heyv84} 
Heyvaerts, J., \& Priest, E.~R.\ 1984, \aap, 137, 63 


\bibitem[Jeong 
\& Chae(2007)]{jeong07} Jeong, H., \& Chae, J.\ 2007, \apj, 671, 1022 

\bibitem[Jing et al.(2010)]{jing10} Jing, J., Tan, C., Yuan, 
Y., Wang, B., Wiegelmann, T., Xu, Y., \& Wang, H.\ 2010, \apj, 713, 440 


\bibitem[Leighton(1964)]{leigh64} Leighton, R.~B.\ 1964, \apj, 140, 1547 

\bibitem[Mackay et al.(2000)]{mac00} 
Mackay, D.~H., Gaizauskas, V., \& van Ballegooijen, A.~A.\ 2000, \apj, 544, 1122 

\bibitem[Mackay \& Gaizauskas(2003)]{mac03} 
Mackay, D.~H., \& Gaizauskas, V.\ 2003, \solphys, 216, 121 

\bibitem[Mackay \& van Ballegooijen(2005)]{mac05} 
Mackay, D.~H., \& van Ballegooijen, A.~A.\ 2005, \apjl, 621, L77

\bibitem[Mackay \& van Ballegooijen(2006a)]{mac06a} 
Mackay, D.~H., \& van Ballegooijen, A.~A.\ 2006, \apj, 641, 577 

\bibitem[Mackay \& van Ballegooijen(2006b)]{mac06b} 
Mackay, D.~H., \& van Ballegooijen, A.~A.\ 2006, \apj, 642, 1193 

\bibitem[Mackay \& van Ballegooijen(2009)]{mac09} 
Mackay, D.~H., \& van Ballegooijen, A.~A.\ 2009, \solphys, 260, 321 

\bibitem[Magara(2004)]{mag04} 
Magara, T.\ 2004, \apj, 605, 480 

\bibitem[Metcalf et al.(2008)]{met08} 
Metcalf, T.~R., et al.2008, \solphys, 247, 269 

\bibitem[Murray \& Hood(2008)]{murr08} 
Murray, M.~J., \& Hood, A.~W.\ 2008, \aap, 479, 567 

\bibitem[Parker(1988)]{Parker88} 
Parker, E.~N.\ 1988, \apj, 330, 474 

\bibitem[Pariat et al.(2005)]{pariat05} 
Pariat, E., D{\'e}moulin, P., \& Berger, M.~A.\ 2005, \aap, 439, 1191

\bibitem[Pevtsov et al.(1995)]{pevtsov95} 
Pevtsov, A.~A., Canfield, R.~C., \& Metcalf, T.~R.\ 1995, \apjl, 440, L109 


\bibitem[Priest(1982)]{priest82} 
Priest, E.~R.\ 1982, Dordrecht, Holland ; Boston : D.~Reidel Pub.~Co.~; Hingham,, 74P 

\bibitem[Regnier \& Priest(2007)]{reg07} 
Regnier, S., \& Priest, E.~R.\ 2007, \apjl, 669, L53 

\bibitem[R{\'e}gnier(2009)]{reg09} R{\'e}gnier, S.\ 2009, \aap, 497, L17 

\bibitem[Savcheva \& van Ballegooijen(2009)]{sav09}
Savcheva, A., \& van Ballegooijen, A. 2009, ApJ, 703, 1766

\bibitem[Scherrer et al.(1995)]{Scherrer95}
Scherrer, P.H., Bogart, R.S., Bush, R.I., Hoeksema, J.T., Kosovichev,
A.G., Schou, J., Rosenburg, W., Springer, L., Tarbell, T.D., {\it{et al.}}:
1995, \solphys, 162, 129

\bibitem[Schrijver et al.(2006)]{sch06} 
Schrijver, C.~J., et  al.\ 2006, \solphys, 235, 161 

\bibitem[Su et al.(2009a)]{su09} 
Su, J.~T., Sakurai, T., Suematsu, Y., Hagino, M., \& Liu, Y.\ 2009, \apjl, 697, L103 

\bibitem[Su et al.(2009b)]{su09b}  
Su, Y., van Ballegooijen, A., Lites, B.W., DeLuca, E.E.,  Golub, L., Grigis, P.C., 
Huang, G., \& Ji, H. 2009, ApJ, 691, 105

\bibitem[Tian \& Alexander(2008)]{tian08} Tian, L., \& Alexander, D.\ 2008, \apj, 673, 532 

\bibitem[van Ballegooijen et al.(2000)]{bal00} 
van Ballegooijen, A.~A., Priest, E.~R., \& Mackay, D.~H.\ 2000, \apj, 539, 983 

\bibitem[van Ballegooijen(2004)]{bal04} 
van Ballegooijen, A.~A.\ 2004, \apj, 612, 519 

\bibitem[Wheatland \& R{\'e}gnier(2009)]{wheat09} 
Wheatland, M.~S., \& R{\'e}gnier, S.\ 2009, \apjl, 700, L88 

\bibitem[Wilmot-Smith et al.(2010)]{Wilmot-Smith10} 
Wilmot-Smith, A.~L., Pontin, D.~I., \& Hornig, G.\ 2010, \aap, 516, A5 

\bibitem[Woltjer(1958)]{wolt58} 
Woltjer, L.\ 1958, Proceedings of the National Academy of Science, 44, 489 

\bibitem[Yeates et al.(2007)]{yea07} 
Yeates, A.~R., Mackay, D.~H., \& van Ballegooijen, A.~A.\ 2007, \solphys, 245, 87 

\bibitem[Yeates et al.(2008)]{yea08} 
Yeates, A.~R., Mackay, D.~H., \& van Ballegooijen, A.~A.\ 2008, \solphys, 247, 103 

\bibitem[Yeates \& Mackay(2009)]{yea09} 
Yeates, A.~R., \& Mackay, D.~H.\ 2009, \apj, 699, 1024 

\bibitem[Yeates et al.(2010)]{yea10a} 
Yeates, A.~R., Attrill, G.~D.~R., Nandy, D., Mackay, D.~H., Martens, P.~C.~H., \& van Ballegooijen, 
A.~A.\ 2010, \apj, 709, 1238 

\bibitem[Yeates et al.(2010)]{yea10b} 
Yeates, A.~R. Mackay, D.~H. van Ballegooijen, A.A., 2010, \jgr, accepted
\end{thebibliography}
\end{document}